\PassOptionsToPackage{unicode}{hyperref}
\PassOptionsToPackage{hyphens}{url}
\PassOptionsToPackage{dvipsnames,svgnames,x11names}{xcolor}
\documentclass[
]{article}

\usepackage{amsmath,amssymb}
\usepackage{iftex}
\ifPDFTeX
  \usepackage[T1]{fontenc}
  \usepackage[utf8]{inputenc}
  \usepackage{textcomp} 
\else 
  \usepackage{unicode-math}
  \defaultfontfeatures{Scale=MatchLowercase}
  \defaultfontfeatures[\rmfamily]{Ligatures=TeX,Scale=1}
\fi
\usepackage{lmodern}
\ifPDFTeX\else  
\fi
\IfFileExists{upquote.sty}{\usepackage{upquote}}{}
\IfFileExists{microtype.sty}{
  \usepackage[]{microtype}
  \UseMicrotypeSet[protrusion]{basicmath} 
}{}
\makeatletter
\@ifundefined{KOMAClassName}{
  \IfFileExists{parskip.sty}{%
    \usepackage{parskip}
  }{
    \setlength{\parindent}{0pt}
    \setlength{\parskip}{6pt plus 2pt minus 1pt}}
}{
  \KOMAoptions{parskip=half}}
\makeatother
\usepackage{xcolor}
\makeatletter
\ifx\paragraph\undefined\else
  \let\oldparagraph\paragraph
  \renewcommand{\paragraph}{
    \@ifstar
      \xxxParagraphStar
      \xxxParagraphNoStar
  }
  \newcommand{\xxxParagraphStar}[1]{\oldparagraph*{#1}\mbox{}}
  \newcommand{\xxxParagraphNoStar}[1]{\oldparagraph{#1}\mbox{}}
\fi
\ifx\subparagraph\undefined\else
  \let\oldsubparagraph\subparagraph
  \renewcommand{\subparagraph}{
    \@ifstar
      \xxxSubParagraphStar
      \xxxSubParagraphNoStar
  }
  \newcommand{\xxxSubParagraphStar}[1]{\oldsubparagraph*{#1}\mbox{}}
  \newcommand{\xxxSubParagraphNoStar}[1]{\oldsubparagraph{#1}\mbox{}}
\fi
\makeatother

\providecommand{\tightlist}{%
  \setlength{\itemsep}{0pt}\setlength{\parskip}{0pt}}\usepackage{longtable,booktabs,array}
\usepackage{calc} 
\usepackage{etoolbox}
\makeatletter
\patchcmd\longtable{\par}{\if@noskipsec\mbox{}\fi\par}{}{}
\makeatother
\IfFileExists{footnotehyper.sty}{\usepackage{footnotehyper}}{\usepackage{footnote}}
\makesavenoteenv{longtable}
\usepackage{graphicx}
\makeatletter
\newsavebox\pandoc@box
\newcommand*\pandocbounded[1]{
  \sbox\pandoc@box{#1}%
  \Gscale@div\@tempa{\textheight}{\dimexpr\ht\pandoc@box+\dp\pandoc@box\relax}%
  \Gscale@div\@tempb{\linewidth}{\wd\pandoc@box}%
  \ifdim\@tempb\p@<\@tempa\p@\let\@tempa\@tempb\fi
  \ifdim\@tempa\p@<\p@\scalebox{\@tempa}{\usebox\pandoc@box}%
  \else\usebox{\pandoc@box}%
  \fi%
}
\def\fps@figure{htbp}
\makeatother
\NewDocumentCommand\citeproctext{}{}

\makeatletter
 \let\@cite@ofmt\@firstofone
 \def\@biblabel#1{}
 \def\@cite#1#2{{#1\if@tempswa , #2\fi}}
\makeatother
\newlength{\cslhangindent}
\newlength{\csllabelwidth}
\newenvironment{CSLReferences}[2] 
 {\begin{list}{}{%
  \setlength{\itemindent}{0pt}
  \setlength{\leftmargin}{0pt}
  \setlength{\parsep}{0pt}
  \ifodd #1
   \setlength{\leftmargin}{\cslhangindent}
   \setlength{\itemindent}{-1\cslhangindent}
  \fi
  \setlength{\itemsep}{#2\baselineskip}}}
 {\end{list}}
\usepackage{calc}

\usepackage{arxiv}
\usepackage{orcidlink}
\usepackage{amsmath}
\usepackage[T1]{fontenc}
\makeatletter
\@ifpackageloaded{caption}{}{\usepackage{caption}}
\AtBeginDocument{%
\ifdefined\contentsname
  \renewcommand*\contentsname{Table of contents}
\else
  \newcommand\contentsname{Table of contents}
\fi
\ifdefined\listfigurename
  \renewcommand*\listfigurename{List of Figures}
\else
  \newcommand\listfigurename{List of Figures}
\fi
\ifdefined\listtablename
  \renewcommand*\listtablename{List of Tables}
\else
  \newcommand\listtablename{List of Tables}
\fi
\ifdefined\figurename
  \renewcommand*\figurename{Figure}
\else
  \newcommand\figurename{Figure}
\fi
\ifdefined\tablename
  \renewcommand*\tablename{Table}
\else
  \newcommand\tablename{Table}
\fi
}
\@ifpackageloaded{float}{}{\usepackage{float}}
\floatstyle{ruled}
\@ifundefined{c@chapter}{\newfloat{codelisting}{h}{lop}}{\newfloat{codelisting}{h}{lop}[chapter]}
\floatname{codelisting}{Listing}

\makeatother
\makeatletter
\@ifpackageloaded{caption}{}{\usepackage{caption}}
\@ifpackageloaded{subcaption}{}{\usepackage{subcaption}}
\makeatother

\usepackage{bookmark}

\IfFileExists{xurl.sty}{\usepackage{xurl}}{} 
\hypersetup{
  pdftitle={PyPotteryInk: One-Step Diffusion Model for Sketch to Publication-ready Archaeological Drawings},
  pdfauthor={Lorenzo Cardarelli},
  pdfkeywords={Pottery, Generative AI, Archaeological
Drawing, Image-to-Image Translation, Diffusion Models},
  colorlinks=true,
  linkcolor={blue},
  filecolor={Maroon},
  citecolor={Blue},
  urlcolor={Blue},
  pdfcreator={LaTeX via pandoc}}

\newcommand{\runninghead}{A Preprint }
\renewcommand{\runninghead}{A Preprint }
\title{PyPotteryInk: One-Step Diffusion Model for Sketch to
Publication-ready Archaeological Drawings}
\author{\textbf{Lorenzo
Cardarelli}~\orcidlink{0000-0002-2436-9967}\\Department of Science of
Antiquities\\Sapienza University of
Rome\\Rome,\ 185\\\href{mailto:lorenzo.cardarelli@uniroma1.it}{lorenzo.cardarelli@uniroma1.it}}
\date{}
\begin{document}
\maketitle
\begin{abstract}
Archaeological pottery documentation traditionally requires a
time-consuming manual process of converting pencil sketches into
publication-ready inked drawings. I present \emph{PyPotteryInk}, an
open-source automated pipeline that transforms archaeological pottery
sketches into standardised publication-ready drawings using a one-step
diffusion model. Built on a modified \emph{img2img-turbo} architecture,
the system processes drawings in a single forward pass while preserving
crucial morphological details and maintaining archaeologic documentation
standards and analytical value. The model employs an efficient
patch-based approach with dynamic overlap, enabling high-resolution
output regardless of input drawing size. I demonstrate the effectiveness
of the approach on a dataset of Italian protohistoric pottery drawings,
where it successfully captures both fine details like decorative
patterns and structural elements like vessel profiles or handling
elements. Expert evaluation confirms that the generated drawings meet
publication standards while significantly reducing processing time from
hours to seconds per drawing. The model can be fine-tuned to adapt to
different archaeological contexts with minimal training data, making it
versatile across various pottery documentation styles. The pre-trained
models, the Python library and comprehensive documentation are provided
to facilitate adoption within the archaeological research community.
\end{abstract}
{\bfseries \emph Keywords}
\def\sep{\textbullet\ }
Pottery \sep Generative AI \sep Archaeological
Drawing \sep Image-to-Image Translation \sep 
Diffusion Models

\section{Introduction}\label{introduction}

Archaeological ceramics are a valuable source of information for
reconstructing the customs, exchanges and social relationships of
ancient populations, as well as for dating archaeological contexts
(Sinopoli 1991; Peroni 1994; Steiner and Allason-Jones 2005; Vidale
2007; Orton and Hughes 2013; Hunt 2016). However, in order to turn a
ceramic fragment into a rich source of scientific information, a long
process of study and elaboration is required: once recovered in an
excavation, the ceramic fragment is washed, catalogued, drawn and made
ready for publication through the preparation of tables and figures that
allow its correct interpretation and comparison with other
archaeological contexts. Archaeological drawing is a fundamental and
well-established tool in archaeological practice, and new technologies
and methods are emerging to automate, standardise and speed up this
process as much as possible. An example of this is the LAD (Laser Aided
Profiler - Demján, Pavúk, and Roosevelt 2023), a tool that allows
ceramic fragments to be `drawn' quickly and accurately using a laser
beam. Over time, however, many drawings were made by hand using
traditional tools such as pencils and then had to be `inked' and made
ready for publication. Traditionally, this post-process was done by hand
with Indian ink, and nowadays digital drawing programmes are used. This
process is however extremely time-consuming and can often discourage the
publication of new contexts due to the difficulties in terms of time and
resources needed for inking.

Generative AI can help to achieve this task, using complex image
translation operation. Today, AI is permeating business, creativity and
everyday life (Elliott 2019; Le et al. 2020; Varghese, Raj, and
Venkatesh 2022; Azatbekova 2023) and its effects are becoming
increasingly apparent in archaeology and cultural heritage (Bickler
2021; Cacciari and Pocobelli 2022; Gattiglia 2025), where it is being
used to classify artefacts (Anichini et al. 2021; Pawlowicz and Downum
2021; Ling et al. 2024), discover hidden structures (Lyons, Fecher, and
Reindel 2022; Kadhim and Abed 2023; Sakai et al. 2023), reconstruct
fragmented data (Navarro et al. 2022; L. Cardarelli 2024b; Altaweel,
Khelifi, and Zafar 2024) and also to analyse data (regarding ceramics:
Navarro et al. 2021; Parisotto et al. 2022; L. Cardarelli 2022, 2024a;
Pang et al. 2024).

This paper proposes \emph{PyPotteryInk}, a Python packages that
leverages deep learning (DL) and traditional image processing pipeline
for automating the digital inking process of ceramic drawings. The model
uses state-of-the-art DL technologies to translate a pencil sketch into
a inked, publication-ready drawing while preserving the original
characteristics of the fragment and enhance their analytical power.

\section{Research Aims}\label{research-aims}

This research addresses the need for efficient and accurate
documentation methods in archaeological pottery studies.
\emph{PyPotteryInk} aims to transform traditional documentation
workflows through application of DL and generative methods, specifically
targeting the labour-intensive process of converting sketches to
publication-ready illustrations. The project pursues six interconnected
objectives:

\begin{enumerate}
\def\labelenumi{\arabic{enumi}.}
\tightlist
\item
  Develop a one-step diffusion model for converting archaeological
  sketches to publication-ready drawings.
\item
  Create an efficient patch-based system for processing drawings of any
  size.
\item
  Enable model adaptation to different pottery styles through
  fine-tuning.
\item
  Validate output quality through expert archaeological assessment.
\item
  Provide an accessible tool for the archaeological community through
  open-source software and documentation.
\item
  Demonstrate significant time savings in the documentation workflow
  while maintaining quality standards.
\end{enumerate}

\section{Materials and Methods}\label{materials-and-methods}

\subsection{Archaeological drawings and digital
inking}\label{archaeological-drawings-and-digital-inking}

Archaeological drawing is a fundamental part of the documentation and
publication of archaeological finds (Griffiths, Jenner, and Wilson 2002;
Steiner and Allason-Jones 2005). It makes it possible to produce a
standardised, two-dimensional representation of an artefact according to
a set of conventions that are followed everywhere, with some minor
variations. Within various traditions, this work fits within the Italian
tradition of protohistory, which favours the following standardised
conventions.

The drawing consists of several components (1) the profile section on
the left end; (2) the exterior view (or prospectus) on the right; (3)
the exterior profile on the right end; (4) a series of lines defining
the axis of symmetry or the diameter line. The profile section reveals
the vessel's internal structure through a cross-sectional cut, including
wall thickness, rims, and any applied features such as handles or
decorative elements. The exterior view displays decorative patterns and
the overall shape of the fragment. The horizontal distance between the
profile and exterior profile corresponds directly to the vessel's
maximum diameter or width. Some scholars also include detailed rendering
of surface textures and finishing treatments in the exterior view.
Traditionally, the drawing is done in pencil and the shadows on the
prospectus are quickly rendered by a play of \emph{chiaroscuro}. When
the artefact is published, the drawing is `inked', i.e.~re-drawn by hand
with Indian ink or using a graphics program such as Adobe Illustrator or
Inkscape. This polishing process results in a cleaner and more defined
drawing, with clean and homogeneous lines, while shadows are represented
by a fine dotting. Inking is an artistic process and requires a certain
amount of time and expertise, and a single drawing can take from minutes
to several hours of work. The inking process is also essential for the
interpretation of the artefact: especially in the case of morphological
analysis (both traditional and digital - L. Cardarelli 2023), the inking
lead to a more standardised result, which makes it easier to compare the
artefact's characteristics between different assemblages.

\subsection{Image-to-image translation and diffusion
models}\label{image-to-image-translation-and-diffusion-models}

The task for the proposed model is to transform a sketch into a
ready-to-publish drawing. In AI, this task fall within image-to-image
translation. Among the first methods proposed to handle this task are
Isola et al. (2018), where the job is tackled with the use of Generative
Adversarial Networks (GANs). GANs are a type of artificial neural
network used to generate new synthetic data from real data.
Specifically, they consist of two neural networks, a generator \(G(x)\)
and a discriminator \(D(x)\), which compete in a zero-sum game
(Goodfellow et al. 2014; Gui et al. 2020). The generator tries to
produce synthetic data that is indistinguishable from real data, while
the discriminator tries to distinguish real data from synthetic data.
GANs have enjoyed considerable success, even in archaeological
applications, where their generative power has been mainly used to
restore fragmented artefacts (Navarro et al. 2022; Altaweel, Khelifi,
and Zafar 2024). However, training a GAN model can be difficult due to
problems such as training instability and mode collapse (Saad, O'Reilly,
and Rehmani 2023).

As of 2021, new Denoising Diffusion Probabilistic Models (DDPMs) have
demonstrated the ability to generate better quality images (Dhariwal and
Nichol 2021) and define the current state-of-the-art for the image
generation task, including commercially used models such as OpenAI's
DALL-E or Midjourney (Ramesh et al. 2021) as well as open-source models
like Stability AI's Stable Diffusion, which will be use as a basis for
the model proposed in this paper. DDPMs have already been used in
archaeological applications: for example X. Zhang (2024), successfully
used a fine-tuned model to reconstruct complex ceramic decoration
patterns, while Jaramillo and Sipiran (2024) applied a DDPM model to
restore point cloud of tridimensional cultural heritage artefacts.

The principle of DDPMs can be conceptualised through a two-phase
process: initially, Gaussian noise is systematically introduced to an
image through a \emph{forward process} comprising multiple sequential
steps, followed by a \emph{reverse process} that progressively removes
this noise to generate the desired output (Sohl-Dickstein et al. 2015).
Stable Diffusion (Rombach et al. 2022) represents an advancement in this
domain by implementing the diffusion process within a latent space
generated by a Variational Autoencoder (VAE - Kingma and Welling 2022),
rather than operating directly in pixel space. This architectural
innovation yields two critical advantages: it substantially reduces
computational requirements and enables the conditioning of the diffusion
process on textual embeddings. Specifically, this allows textual prompts
to guide the generative process, providing precise control over the
output characteristics. While Stable Diffusion excels at text-to-image
generation, archaeological documentation requirements necessitate
image-to-image translation capabilities. To address this methodological
gap, an adaptation of Stable Diffusion optimized for image-to-image
translation tasks called \emph{img2img-turbo} is used (Parmar et al.
2024). The authors propose an implementation that works both for paired
(\emph{pix2pix-turbo}) and unpaired (\emph{cycleGAN-turbo}) image
translation tasks, with the first being considered the most suitable for
purpose of this work.

Here is a brief overview of the model architecture and training process:

\begin{enumerate}
\def\labelenumi{\arabic{enumi}.}
\item
  \textbf{Input Encoding}: An input image \(x\) is encoded into a
  lower-dimensional latent representation: \[z_{in} = \mathcal{E}(x)\]

  where \(x \in \mathbb{R}^{H \times W \times 3}\) is the input image
  and \(\mathcal{E}\) is the VAE encoder augmented with LoRA adapters
  (\emph{see below}).
\item
  \textbf{\emph{Latent Space Processing}}: Unlike traditional diffusion
  models that require multiple denoising iterations (Sohl-Dickstein et
  al. 2015), Parmar et al. (2024) propose modified U-Net (Ronneberger,
  Fischer, and Brox 2015) that performs the entire denoising process in
  a single forward pass: \[z_{out} = \mathcal{U}(z_{in}, c)\]

  where \(c\) represents the text embedding condition, processed with
  CLIP's text encoder (Radford et al. 2021) (see below). \(\mathcal{U}\)
  denotes the U-Net architecture enhanced with targeted modifications
  for efficient single-step processing.
\item
  \textbf{Decoding}: The denoised latent representation is transformed
  back to image space: \[\hat{x} = \mathcal{D}(z_{out})\]

  where \(\mathcal{D}\) represents the VAE decoder with skip connections
  for preserving high-detail fidelity.
\end{enumerate}

Parmar et al. (2024) propose several losses for training the model
paired translation task:

\begin{enumerate}
\def\labelenumi{\arabic{enumi}.}
\item
  Reconstruction Loss (\(\mathcal{L}_{rec}\)):
  \[\mathcal{L}_{rec} = ||x - \hat{x}||_2 + \lambda_{lpips}L_{lpips}(x, \hat{x})\]

  where \(||x - \hat{x}||_2\) is the L2 distance (Euclidean distance)
  between \(x\): the target (ground truth) image and the generated image
  \(\hat{x}\); \(L_{lpips}\) is the LPIPS perceptual loss that measures
  similarity in feature space and \(\lambda_{lpips}\) is a weight
  parameter to balance the two terms. The L2 term ensures pixel-level
  accuracy, while LPIPS ensures perceptual similarity. LPIPS uses a
  pretrained neural network (typically VGG - Simonyan and Zisserman
  2015) to compare images in feature space rather than pixel space,
  which better matches human perception of image similarity (R. Zhang et
  al. 2018).
\item
  Adversarial Loss (\(\mathcal{L}_{GAN}\)):
  \[\mathcal{L}_{GAN} = \mathbb{E}_{x}[\log D(x)] + \mathbb{E}_{x}[\log(1 - D(\hat{x}))]\]

  This is the standard GAN loss, where discriminator (\(D\)) tries to
  distinguish between real images \(x\) and generated images
  \(\hat{x}\), while the generator (\(G\)) tries to fool the
  discriminator.
\end{enumerate}

In conclusion, the training objective is to minimise the reconstruction
loss \(\mathcal{L}_{rec}\), the adversarial loss \(\mathcal{L}_{GAN}\),
and the CLIP text-image alignment loss \(\mathcal{L}_{CLIP}\):

\[\mathcal{L} = \mathcal{L}_{rec} + \lambda_{GAN}\mathcal{L}_{GAN} + \lambda_{CLIP}\mathcal{L}_{CLIP}\]

where \(\lambda_{GAN}\) and \(\lambda_{CLIP}\) are hyperparameters that
balance the different losses. For this implementation, all training
parameters are available in the project repository.

Although the model accepts a textual prompt as a condition, this is
irrelevant for our purposes, as the need is to produce a single style of
output. Therefore, the model is trained with a fixed prompt
(``\emph{make it ready for publication}'') to ensure a consistent output
style.

During the encoding process, the model uses LoRA (Low-Rank Adaptation)
to reduce the number of parameters and computational overhead (Hu et al.
2021). In other words, LoRA allows the model to adapt to new tasks with
minimal additional parameters, making it ideal for fine-tuning on new
examples. The key innovation of LoRA lies in decomposing the weight
updates into low-rank matrices:

Instead of updating the full weight matrix \(W\), LoRA decomposes the
update into: \[\Delta W = BA\]

where \(B \in \mathbb{R}^{d \times r}\) and
\(A \in \mathbb{R}^{r \times k}\), \(r\) is the adaptation rank
(typically much smaller than \(d\) and \(k\)) while the original weight
matrix \(W \in \mathbb{R}^{d \times k}\) remains frozen

The capacity of the model to adapt to new tasks is considered crucial
for archaeological applications, where the model must be able to learn
from a small dataset and then be fine-tuned on new contexts or styles.

\subsection{Archaeological dataset used and general training
process}\label{sec-dataset}

As a paired image translation task, a pencil sketch and an inked version
of the same drawing are required to train the model. Simplifying, the
model can learn the relationship between the two styles (or domains) and
reproduce it in inference. The dataset therefore consists of pairs of
492 pots from the Casinalbo (Andrea Cardarelli 2014), Montale (Andrea
Cardarelli 2009), Monte Croce Guardia (Andrea Cardarelli et al. 2017)
and Monte Cimino (Barbaro et al. 2011; A. Cardarelli and Trucco 2014)
contexts. Casinalbo and Montale are MBA-RBA contexts, while Monte Croce
Guardia and Monte Cimino are FBA contexts\footnote{MBA (Middle Bronze
  Age): 1650 - 1325 BCE; RBA (Recent Bronze Age): 1325 - 1150 BCE; FBA
  (Final Bronze Age): 1150 - 925 BCE. (Andrea Cardarelli 2018, 360).}.
The drawings were provided by chair of European Protohistory, Sapienza
University of Rome and Museo Civico di Modena
(Section~\ref{sec-acknowledgements}).

\begin{figure}

\centering{

\pandocbounded{\includegraphics[keepaspectratio]{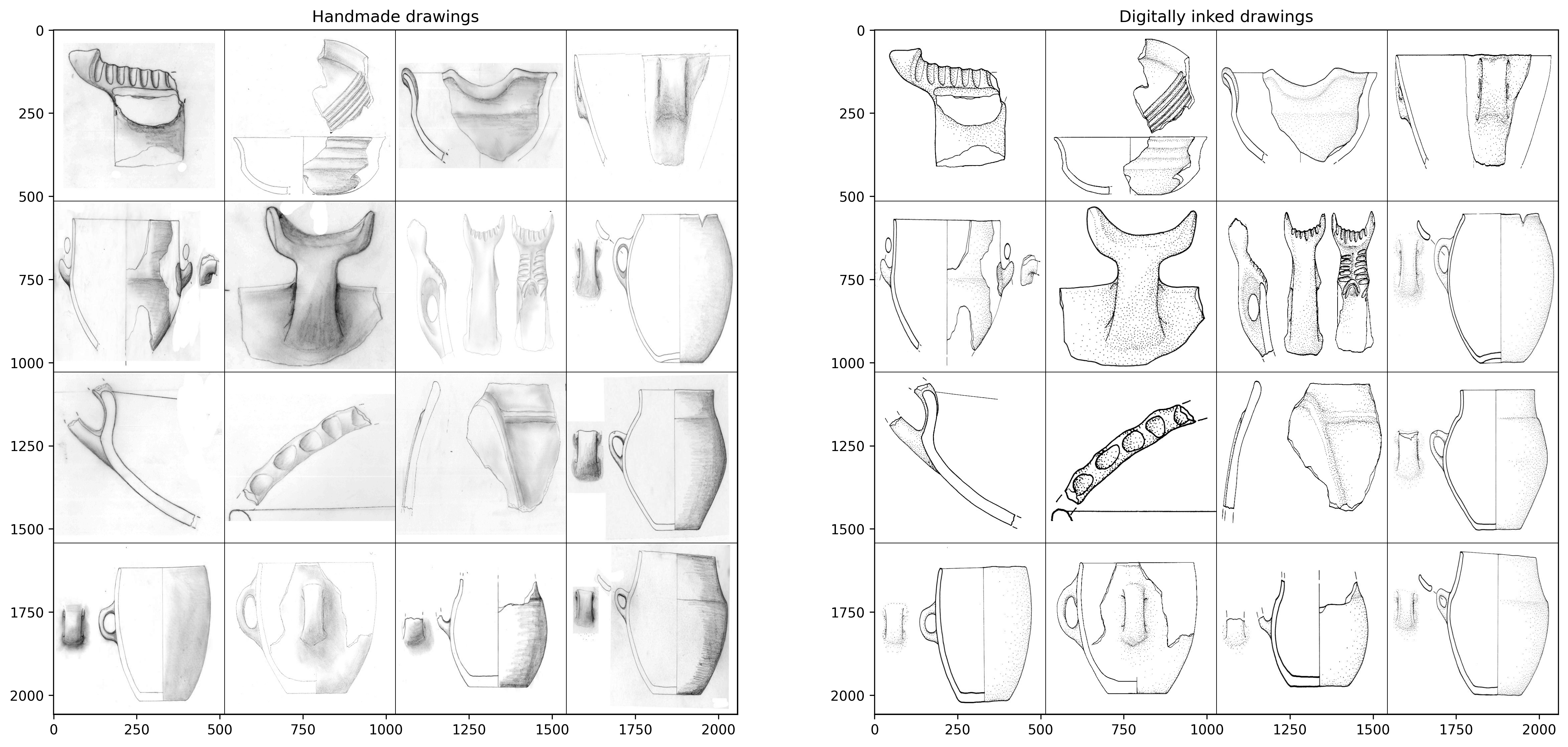}}

}

\caption{\label{fig-10kbatch}Training dataset for the `\emph{10k}'
model. Images are ``compressed'' into a squared box for training
purpose}

\end{figure}

The pottery assemblage used in this experiment consists of protohistoric
vessels manufactured using the characteristic \emph{impasto} technique
of the period. These vessels are handmade with coarse-tempered clay
fired to produce distinctive brownish surfaces
(Figure~\ref{fig-10kbatch}). The decorative elements include both
incised geometric patterns and applied clay elements such as cordons and
lugs, representing typical stylistic features of Italian north-central
protohistoric pottery production (Levi 2010, 194--200).

To train the model, a dataset was used in which the drawings were scaled
and resized within a square of \(512 \cdot 512\) pixels, without any
augmentation (Figure~\ref{fig-10kbatch}). The main objective of this
phase is to create a solid model suitable for general use (called the
`\emph{10k}' model - \emph{see below}), which can then be adapted to
specific contexts through fine-tuning. The dataset was divided into a
training set of 440 images and a validation set of 52 images. The
training last for 10,000 steps. While all training parameters are
available in the code repository, the training dataset is not included
due to copyright restrictions as most of the drawings are unpublished.

\subsection{Fine-tuning and inference}\label{fine-tuning-and-inference}

\begin{figure}

\centering{

\pandocbounded{\includegraphics[keepaspectratio]{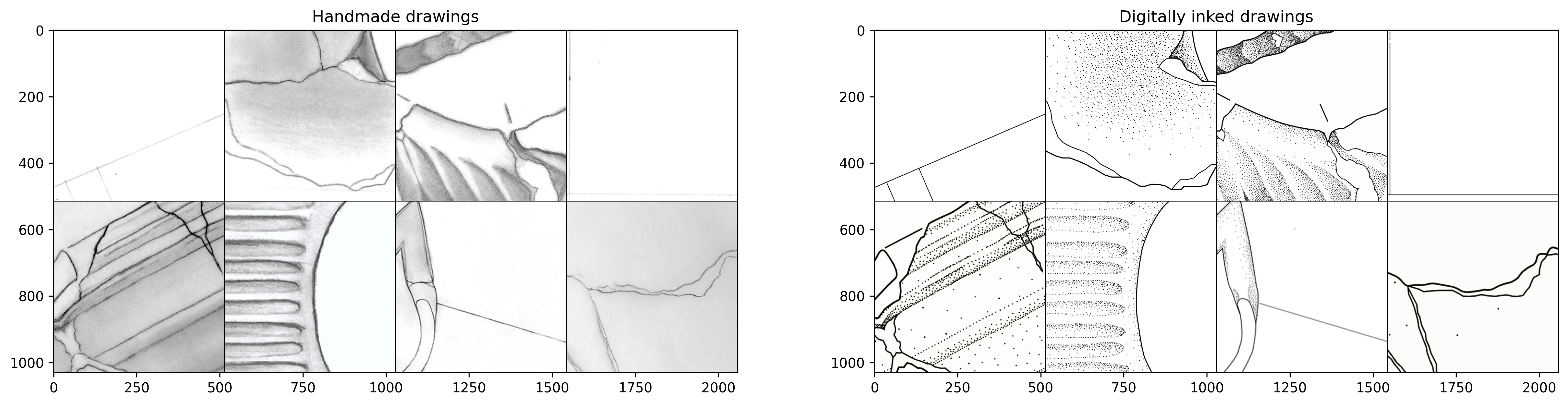}}

}

\caption{\label{fig-6hbatch}Training dataset for the `\emph{6h-MCG}'
model. Extensive data augmentation is used}

\end{figure}

To ensure optimal resolution and maintain the high fidelity required for
archaeological documentation, a patch-based processing methodology is
then implemented. This approach divides each drawing into
\(512 \cdot 512\) pixel segments, allowing the model to capture
fine-grained details. The \emph{10k} model was specifically fine-tuned
on these high-resolution patches to preserve essential archaeological
features such as shadows, decorative patterns, surface treatments, and
precise vessel profiles. 9 pairs of drawings from the Monte Croce
Guardia site were used as fine-tuning training dataset. To compensate
for the limited dataset size, extensive data augmentation is implemented
including the creation of random patches from the source images and
applying rotations, translations, and reflections
(Figure~\ref{fig-6hbatch}). The augmentation strategy effectively
expanded the small dataset, while also testing the model's ability to
adapt to specific archaeological contexts with limited examples. The
training lasts for 600 steps, resulting in the fine-tuned model called
`\emph{6h-MCG}'.

This fine-tuning procedure preludes the inference process, as
illustrated in the figure below (Figure~\ref{fig-patchdivision}):

\begin{figure}

\centering{

\pandocbounded{\includegraphics[keepaspectratio]{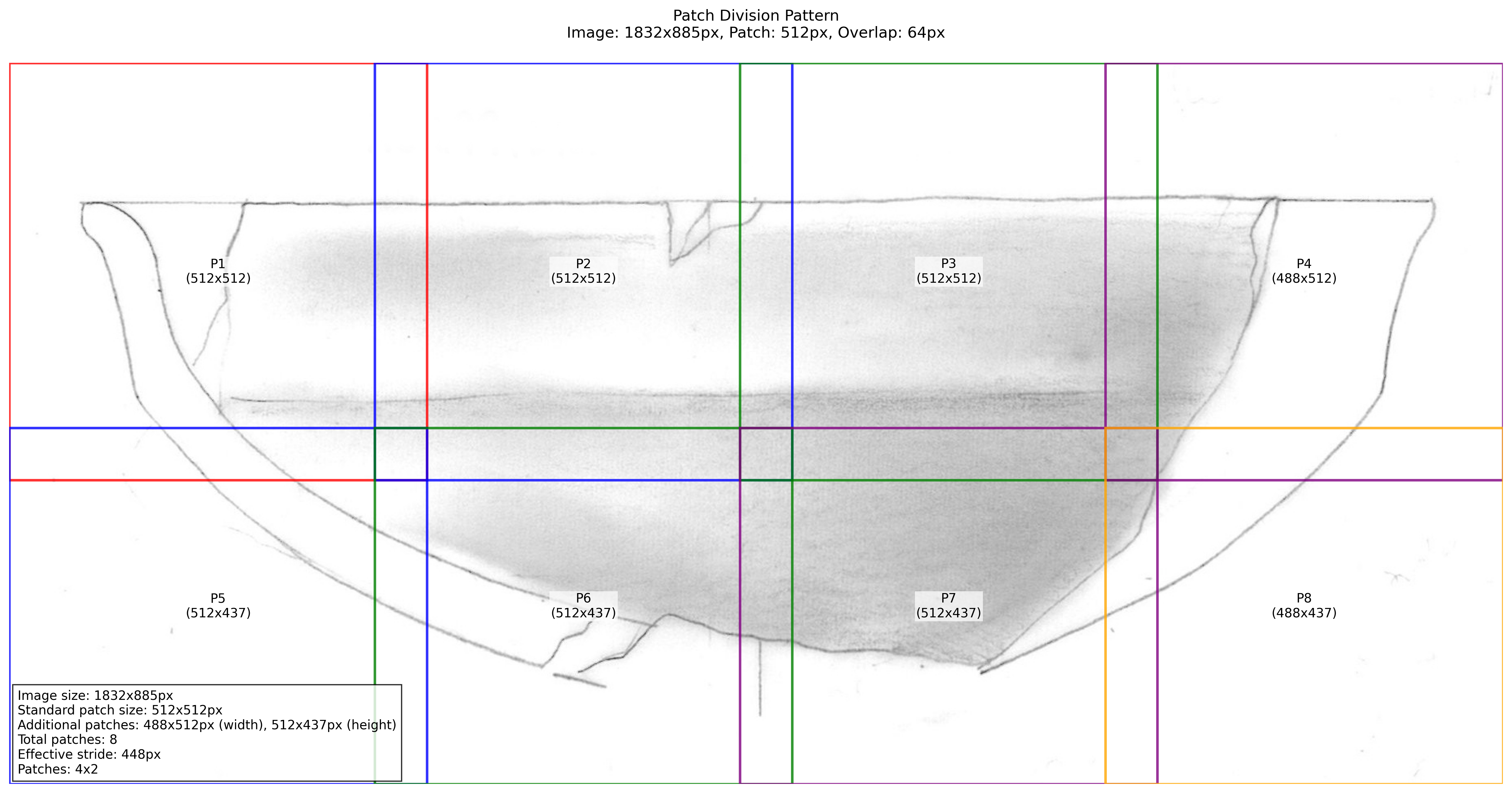}}

}

\caption{\label{fig-patchdivision}Inference patching for an example
image.}

\end{figure}%

In this example, the original \(1832 \cdot 885\) pixels image is divided
into 8 patches (\(512 \cdot 512\) as standard size, plus additional
non-standard patches to fill in gaps) that are processed one at a time,
representing the native resolution of the fine-tuned `\emph{6h-MCG}'
model. The inference algorithm thus ensures that drawings of any size
can be processed maintaining high-quality output. There is also a
64-pixel overlap between the different patches, ensuring a smooth
transition between the different parts of the image and avoiding
splitting artefacts.

In summary, the inference process is as follows:

\begin{enumerate}
\def\labelenumi{\arabic{enumi}.}
\tightlist
\item
  The image is divided into patches of \(512 \cdot 512\) pixels.
\item
  The remaining space is filled with non-standard patches.
\item
  Each patch is processed by the model.
\item
  The results are recombined into a single high-resolution image,
  restoring the original proportions of the image.
\end{enumerate}

\subsection{\texorpdfstring{Code Repository and the \emph{PyPotteryInk}
package}{Code Repository and the PyPotteryInk package}}\label{code-repository-and-the-pypotteryink-package}

This research introduces a modified implementation of the
\emph{img2img-turbo} model adapted for archaeological documentation
purposes. The implementation incorporates several enhancements,
particularly the integration of checkpoint-based training resumption
through LoRA configuration management. This is fundamental for the
fine-tuning process and further development, as a previously trained
models can be used as a starting point. Unfortunately, an initial
explorations of CPU compatibility revealed significant performance
constraints, establishing GPU acceleration as a prerequisite for the
application of the model. The development also focuses on delivering
\emph{PyPotteryInk} as a Python library to facilitate adoption within
the archaeological research community. The package implements
ready-to-use functions for model application, incorporating diagnostic
tools, batch processing capabilities and post-processing steps. The
complete codebase is accessible through the project's GitHub repository
(https://github.com/lrncrd/PyPotteryInk), with pre-trained models
available via Hugging Face, while the documentation is hosted on a
dedicated webpage (https://lrncrd.github.io/PyPotteryInk/)

\section{Results}\label{results}

The analysis of the results was carried out in two stages: the first was
an evaluation of the training metrics and outputs of the model
(Section~\ref{sec-training-results}), while the second was a qualitative
assessment of the results obtained through a comparison of expert
archaeological judgements (Section~\ref{sec-expert-validation}).

\subsection{Training results}\label{sec-training-results}

The various training metrics for the `\emph{10k}' model are proposed in
the following figures (Figure~\ref{fig-trainingmetrics},
Figure~\ref{fig-validationmetrics}):

\begin{figure}

\centering{

\pandocbounded{\includegraphics[keepaspectratio]{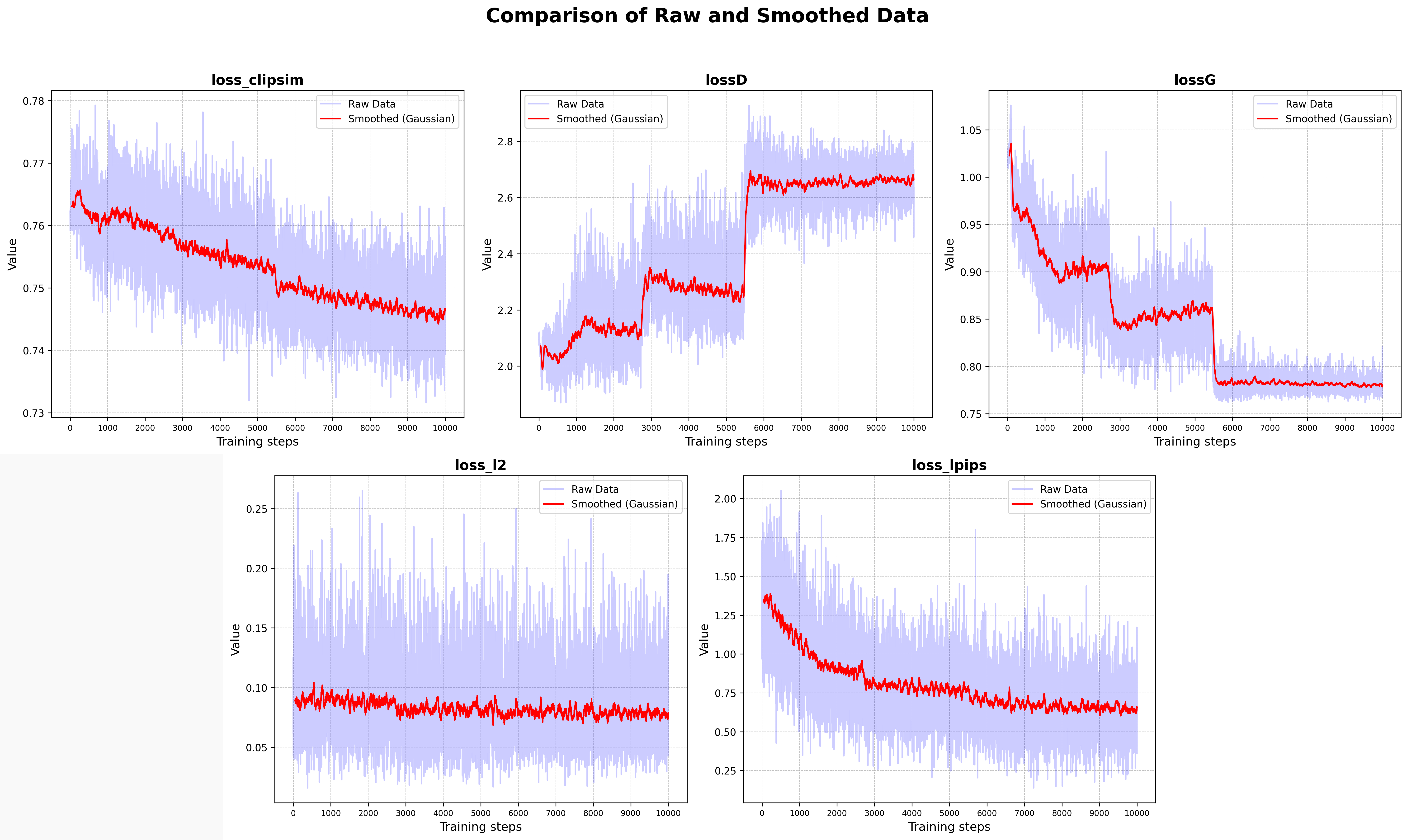}}

}

\caption{\label{fig-trainingmetrics}Training metrics for the \emph{10k}
model.}

\end{figure}%

\begin{figure}

\centering{

\pandocbounded{\includegraphics[keepaspectratio]{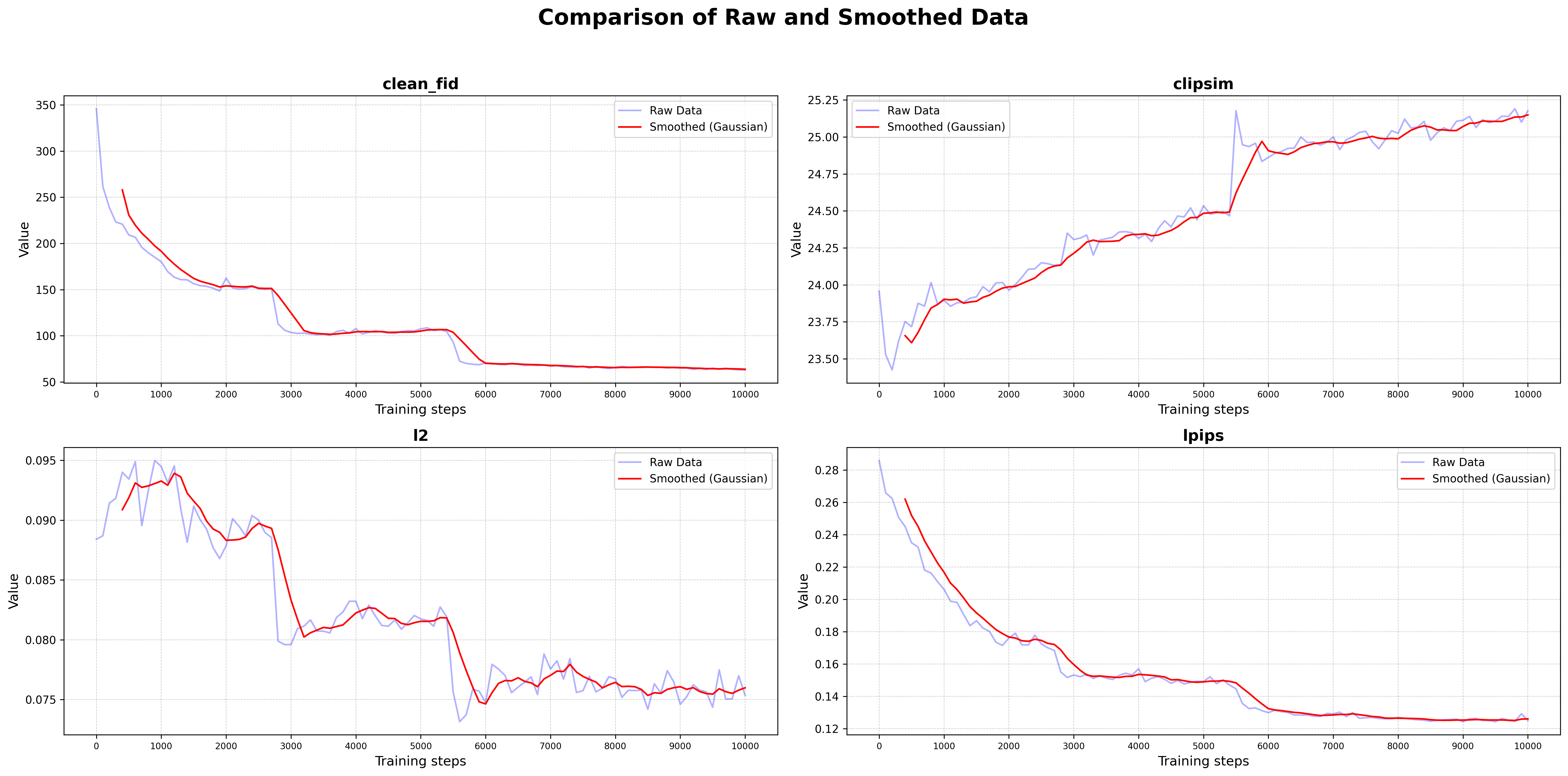}}

}

\caption{\label{fig-validationmetrics}Validation metrics for the
\emph{10k} model}

\end{figure}%

Analysis of the training metrics demonstrates successful model
convergence and learning progression. The discriminator loss (lossD)
exhibits three distinct learning phases before stabilizing around 2.6,
indicating the model successfully learned to distinguish between real
and generated drawings. This is complemented by the generator loss
(lossG) showing stepped decreases before settling at approximately 0.78,
reflecting the model's improving ability to produce convincing
archaeological drawings. The perceptual quality metrics show
particularly encouraging results. The LPIPS perceptual loss demonstrates
substantial improvement, decreasing from 1.25 to 0.65, indicating the
model learned to generate drawings that better match human perception of
visual similarity. This improvement is further validated by the
clean-FID score (Parmar, Zhang, and Zhu 2022), which showed dramatic
enhancement from 350 to 60, confirming increased fidelity between
generated drawings and their targets. While the model tracks several
text-related metrics like CLIP similarity (improving from 23.5 to 25.0),
these are less relevant for our archaeological documentation purpose
since we employ a fixed prompt. Instead, the most significant indicators
are the clean-FID and LPIPS scores, as they directly measure the visual
quality and accuracy of the generated drawings. The model appears to
reach convergence around step 8000, with the stepped nature of the loss
curves reflecting discrete improvements in generation quality. Notably,
these improvements occurred without implementing learning rate decay.

The visual comparison of the model's output during training within the
original sketch and ink versions is also proposed, making it easier to
understand the evolution of the model (Figure~\ref{fig-fidimages}):

\begin{figure}

\centering{

\pandocbounded{\includegraphics[keepaspectratio]{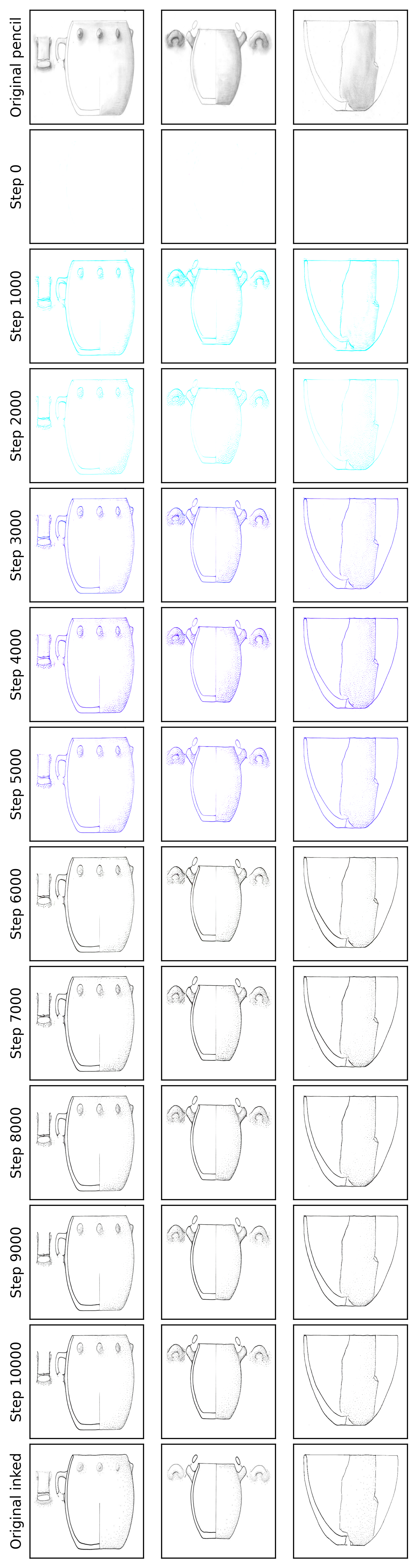}}

}

\caption{\label{fig-fidimages}\emph{10k} model's output during the
training.}

\end{figure}%

While the first epoch doesn't produce any results, a macro-division in 3
phases is evident: (1) Steps 1000-3000: Shows light cyan coloured,
somewhat noisy or sketchy lines (2) Steps 3000-5000: Transitions to
purple-tinted lines, starting to stabilise the drawing (3) Steps
5000-10000: Gradually converges to clean, black line drawings. Double
lines and noise are gradually eliminated, and artifacts are reduced.

These phases relate to the steps in the training metrics
(Figure~\ref{fig-trainingmetrics}), which show a stepped trend that
corresponds to a change in colour and quality of the drawing. In fact,
it should be noted that the original implementation of Stable Diffusion
works with RGB images, whereas the model is adapted to obtain a
greyscale result.

Moving on to the evaluation of the fine-tuned `\emph{6h-MCG}' model, we
can analyse the model's output results within a single example patch
(Figure~\ref{fig-fidimagesval}):

\begin{figure}

\centering{

\pandocbounded{\includegraphics[keepaspectratio]{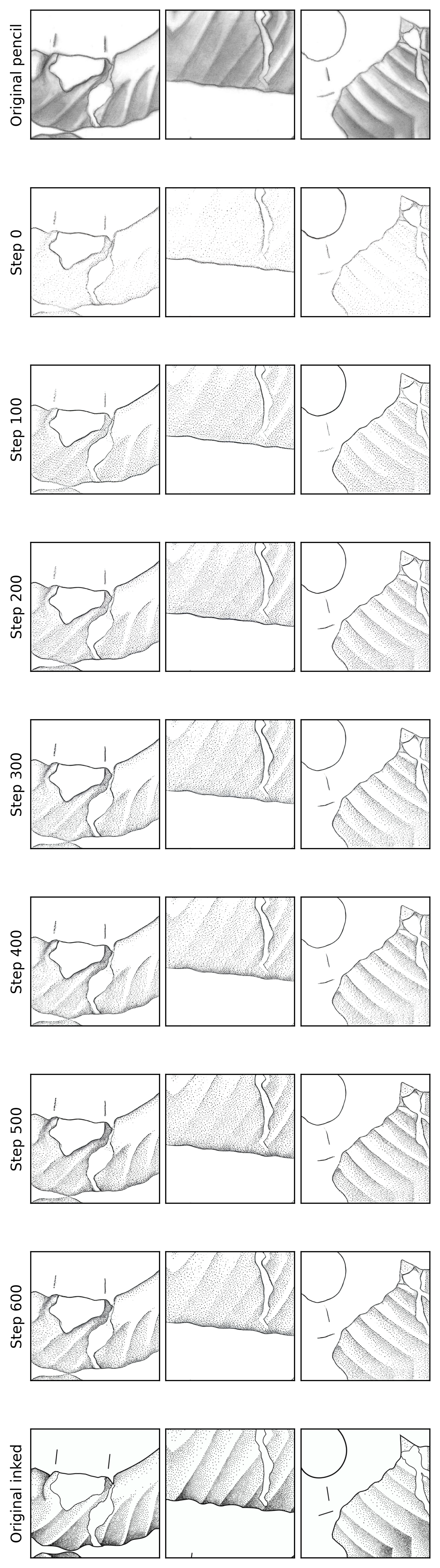}}

}

\caption{\label{fig-fidimagesval}\emph{6h-MCG} model's output.}

\end{figure}%

A clear trend in the evolution of the model is evident: while the first
steps (\textless100) show results with minimal dotting texture, the
model gradually stabilises the shading density and pattern, resulting in
a clean and consistent stippling for shaded areas while maintaining
clear line work.

\subsection{The case study: Montale
pottery}\label{the-case-study-montale-pottery}

\begin{figure}

\centering{

\pandocbounded{\includegraphics[keepaspectratio]{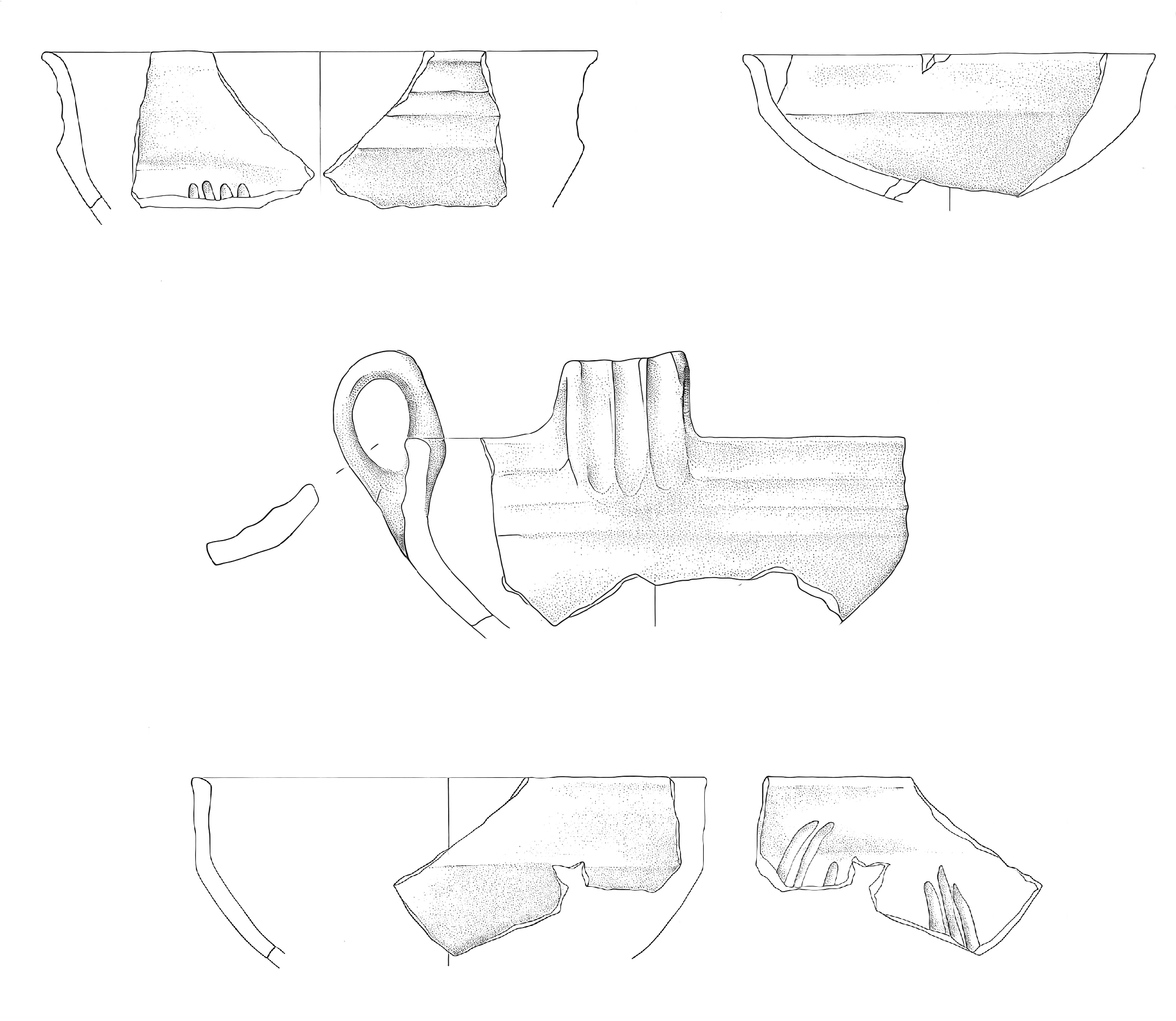}}

}

\caption{\label{fig-table}Some of the AI-inked drawings from Montale
assemblage}

\end{figure}

To test the model's performance, a case study is conducted using ceramic
assemblages from the Montale site (Section~\ref{sec-dataset}). The test
dataset consisted of 72 previously unseen pencil sketches of ceramic
vessels, deliberately excluded from the training corpus to ensure the
model's generalisation capabilities. Some results are shown in
Figure~\ref{fig-table}.

\subsubsection{Expert validation}\label{sec-expert-validation}

If the visual analysis of the output seems to indicate a positive
outcome for the training of the model, it is necessary to compare the
results with the \emph{knowledge-domain} of experts in the field. This
is a crucial step in the validation of the model, as it allows to
understand if the model is able to produce results that are consistent
with the archaeological standards and the analytical requirements of the
field. The results are based on two steps. Firstly, a single-blind
discrimination test is carried out. In this test, a group of
archaeological experts were tasked with distinguishing between
traditionally inked illustrations and model-generated outputs. Secondly,
some key technical attributes are qualitatively evaluated based on a
visual comparison between the original pencil drawings and the model
output.

\paragraph{Single-Blind Discrimination
Test}\label{single-blind-discrimination-test}

A single-blind discrimination test is conducted where four
archaeological experts evaluated a mixed set of traditional and
AI-generated drawings. The experts were aware of the test's purpose but
were blind to the origin of each individual drawing. Their ability to
correctly discriminate between AI-generated and traditional drawings was
measured using accuracy, recall, precision, and Jaccard metrics. To
replicate authentic publication conditions, all test images were scaled
to 1:4 or 1:3 ratio, corresponding to standard publication dimensions in
archaeological literature.

The results of the blind test are shown Table~\ref{tbl-first} and
discussed in the Section~\ref{sec-discussion}:

\begin{longtable}[]{@{}
  >{\raggedright\arraybackslash}p{(\linewidth - 8\tabcolsep) * \real{0.2203}}
  >{\raggedleft\arraybackslash}p{(\linewidth - 8\tabcolsep) * \real{0.2034}}
  >{\raggedleft\arraybackslash}p{(\linewidth - 8\tabcolsep) * \real{0.1695}}
  >{\raggedleft\arraybackslash}p{(\linewidth - 8\tabcolsep) * \real{0.2203}}
  >{\raggedleft\arraybackslash}p{(\linewidth - 8\tabcolsep) * \real{0.1864}}@{}}
\caption{The table offers the results of the single-blind
test.}\label{tbl-first}\tabularnewline
\toprule\noalign{}
\begin{minipage}[b]{\linewidth}\raggedright
Evaluator
\end{minipage} & \begin{minipage}[b]{\linewidth}\raggedleft
Accuracy\footnote{Defines as
  \(Accuracy = \frac{TP + TN}{TP + TN + FP + FN}\), where TP = True
  Positive, TN = True Negative, FP = False Positive, FN = False
  Negative. It rapresents the proportion of correct identifications of
  the model output.}
\end{minipage} & \begin{minipage}[b]{\linewidth}\raggedleft
Recall\footnote{Defides as \(Recall = \frac{TP}{TP + FN}\), where TP =
  True Positive, FN = False Negative. It rapresents the proportion of
  correct identifications of the model output among the total of correct
  identifications.}
\end{minipage} & \begin{minipage}[b]{\linewidth}\raggedleft
Precision\footnote{Defines as \(Precision = \frac{TP}{TP + FP}\), where
  TP = True Positive, FP = False Positive. It rapresents the proportion
  of correct identifications of the model output among the total of
  identifications.}
\end{minipage} & \begin{minipage}[b]{\linewidth}\raggedleft
Jaccard \footnote{Defines as \(Jaccard = \frac{TP}{TP + FP + FN}\),
  where TP = True Positive, FP = False Positive, FN = False Negative. It
  rapresents the proportion of correct identifications of the model
  output among the total of identifications, considering the
  intersection between the two sets.}
\end{minipage} \\
\midrule\noalign{}
\endfirsthead
\toprule\noalign{}
\begin{minipage}[b]{\linewidth}\raggedright
Evaluator
\end{minipage} & \begin{minipage}[b]{\linewidth}\raggedleft
Accuracy{}
\end{minipage} & \begin{minipage}[b]{\linewidth}\raggedleft
Recall{}
\end{minipage} & \begin{minipage}[b]{\linewidth}\raggedleft
Precision{}
\end{minipage} & \begin{minipage}[b]{\linewidth}\raggedleft
Jaccard {}
\end{minipage} \\
\midrule\noalign{}
\endhead
\bottomrule\noalign{}
\endlastfoot
FE & 0.85 & 0.71 & 1 & 0.71 \\
EF & 0.48 & 0.71 & 0.5 & 0.42 \\
LP & 0.32 & 0.26 & 0.31 & 0.16 \\
EP1 & 0.05 & 0.06 & 0.07 & 0.03 \\
\end{longtable}

\paragraph{Qualitative evaluation}\label{qualitative-evaluation}

The qualitative evaluation of the model's output is based on a visual
comparison between the original pencil drawings and the model output.
The experts were asked to evaluate the quality of the output in terms
the following criteria:

\begin{enumerate}
\def\labelenumi{\arabic{enumi}.}
\tightlist
\item
  \textbf{Archaeological consistency} (AC): Does the output respect the
  original characteristics of the artefact?
\item
  \textbf{Line quality} (LQ): Are the lines clean and homogeneous?
\item
  \textbf{Shading quality} (SQ): Is the shading consistent and
  appropriate?
\item
  \textbf{Overall quality} (OQ): Does the output look like a
  publication-ready drawing?
\item
  \textbf{Features recognizability} (FR): Are the features (handles,
  decorations) of the artefact clearly represented?
\item
  \textbf{Further details} (FD): Is the drawing complete and ready for
  publication without requiring manual additions?
\end{enumerate}

The experts were asked to rate each criterion on a scale from 1 to 5,
where:

\begin{itemize}
\tightlist
\item
  1: Very poor or insufficient: the output does not meet the criterion.
\item
  2: Poor: the output partially meets the criterion.
\item
  3: Satisfactory: the output meets the criterion.
\item
  4: Good: the output exceeds the criterion.
\item
  5: Excellent: the output is perfect.
\end{itemize}

The results of the evaluation are shown in Table~\ref{tbl-second} as the
average score.

\begin{longtable}[]{@{}lrrrrrr@{}}
\caption{The table shows the qualitative evaluation
results.}\label{tbl-second}\tabularnewline
\toprule\noalign{}
& AC & LQ & SQ & OQ & FR & FD \\
\midrule\noalign{}
\endfirsthead
\toprule\noalign{}
& AC & LQ & SQ & OQ & FR & FD \\
\midrule\noalign{}
\endhead
\bottomrule\noalign{}
\endlastfoot
CP & 4.94 & 4.87 & 3.51 & 3.86 & 3.53 & 2.74 \\
EP2 & 4.27 & 3.41 & 3.78 & 3.73 & 4.04 & 3.82 \\
ADR & 4.98 & 4.93 & 4.74 & 5 & 4.88 & 4,98 \\
AC & 4.96 & 4.17 & 4.36 & 4.30 & 4.38 & 4.01 \\
\end{longtable}

\subsection{Hardware and software requirements,
scalability}\label{hardware-and-software-requirements-scalability}

Each experiment was performed in a Python (3.10) environment. A NVIDIA
L4 GPU with 24 GB VRAM was used to train the model (\emph{10k} and
\emph{6h-MCG}). You can find the training parameters in the GitHub
project repository. The use of a GPU is mandatory for processing: the
inference process is performed using a RTX 3070Ti with 8 GB VRAM.
Regarding scalability, it is pointless to hide that this process is
highly computational expansive: benchmarks in the project repository
that show the processing time of a test image, so that the applicability
of the model to different hardware can be assessed. For further
considerations on the scalability and limitations of the model, please
refer to the discussion.

\section{Discussion}\label{sec-discussion}

\subsection{Training results and expert
validation}\label{training-results-and-expert-validation}

The results of our single-blind discrimination test present an
intriguing pattern that deserves careful analysis
(Table~\ref{tbl-first}). Most notably, two highly experienced
archaeological illustrators (FE and EP1) showed drastically different
abilities to distinguish AI-generated drawings from traditional ones,
with accuracy scores of 0.85 and 0.05 respectively. FE's high accuracy
and perfect precision (1.0) demonstrate they consistently identified
distinguishing characteristics in AI-generated drawings. In contrast,
EP1's extremely low accuracy (0.05), well below random chance, reveals
they consistently misclassified AI-generated drawings as traditional
ones. The intermediate scores of EF (0.48) and LP (0.32) suggest varying
levels of ability to detect AI-generated content. The dramatic variance
in expert performance highlights the inherent subjectivity in evaluating
archaeological drawings. Different experts prioritise different aspects
of the drawings and apply varying criteria for quality assessment. This
subjectivity is particularly evident in EP1's evaluations - while they
correctly identified many AI-generated drawings, they classified them as
traditional due to their perceived high quality and cleanliness (EP1 -
personal communication). This suggests that preconceptions about AI
capabilities might lead experts to attribute high-quality outputs to
human craftsmanship, rather than artificial intelligence.

The qualitative evaluation scores provide additional context to
understand the model's performance (Table~\ref{tbl-second}).
Archaeological consistency (AC) received high ratings across all experts
(CP: 4.94, EP2: 4.27, ADR: 4.98, AC: 4.96), strongly indicating the
model's reliability in preserving crucial archaeological information.
Line Quality (LQ) shows more variation among experts (CP: 4.87, EP2:
3.41, ADR: 4.93, AC: 4.17), reflecting different professional standards
and expectations in archaeological illustration. The assessment of
Shading Quality (SQ) also varies considerably (CP: 3.51, EP2: 3.78, ADR:
4.74, AC: 4.36), suggesting that the rendering of shadows and textures
is interpreted differently by various experts. Overall Quality (OQ)
ratings span from satisfactory to excellent (CP: 3.86, EP2: 3.73, ADR:
5.00, AC: 4.30), while Feature Recognizability (FR) maintains
consistently good scores (CP: 3.53, EP2: 4.04, ADR: 4.88, AC: 4.38). The
evaluation of Further Details (FD) shows the widest range of scores (CP:
2.74, EP2: 3.82, ADR: 4.98, AC: 4.01), suggesting varying perspectives
on the need for manual intervention or enhancement. These scores -
combined with the discrimination test results - suggest two major
findings. Firstly, the model produces drawings of consistently high
quality that meet archaeological standards, as evidenced by the strong
AC scores across all experts and the difficulty most evaluators had in
distinguishing between AI-generated and traditional drawings in the
single-blind test. Secondly, the evaluation reveals significant
professional subjectivity in assessing archaeological illustrations,
with experts often providing notably different scores for the same
criteria, particularly in areas like Line Quality and Shading Quality.
In terms of practical benefits, the model's ability to produce drawings
in seconds rather than the hours required for manual inking (as noted by
CP and EP2) represents a significant advancement in archaeological
documentation efficiency, while still allowing for manual refinement
when needed.

Moving to reliability, as shown by the \emph{6h-MCG} example, the model
can be fine-tuned on a small dataset to adapt to specific archaeological
contexts. I cannot define a fixed or minimum number of examples needed
for fine-tuning, as this depends on the complexity of the style and the
variability of the new dataset, as well as the differences between the
pottery's styles or morphologies. If the target dataset is so different
in terms of style and morphology, maybe more training examples are
required. As a general rule of thumb, I suggest using at least 10-20
examples for fine-tuning, trying to include as much variability
(especially decorations) as possible.

\subsection{Limitations}\label{limitations}

\begin{figure}

\centering{

\pandocbounded{\includegraphics[keepaspectratio]{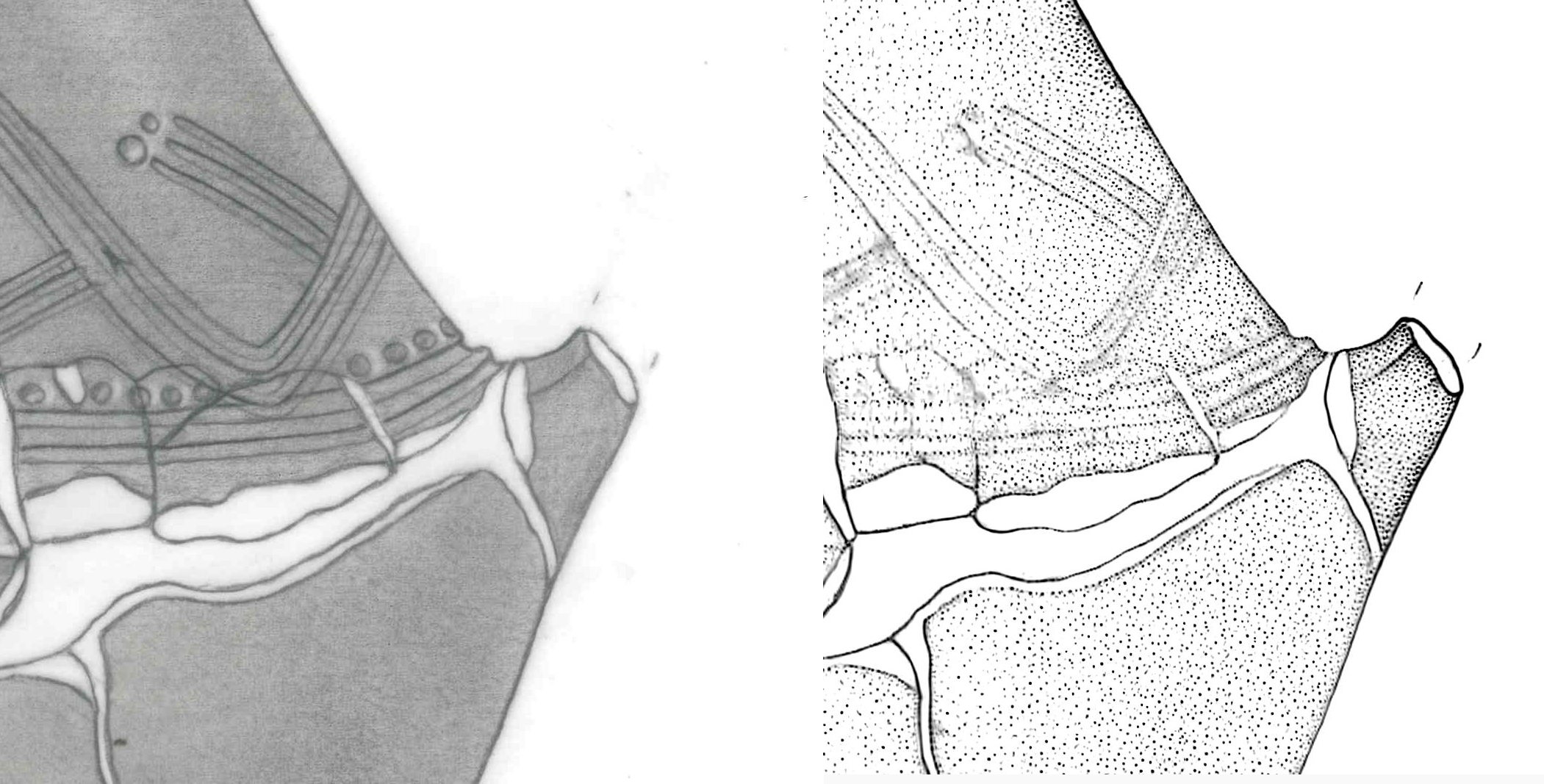}}

}

\caption{\label{fig-limit_1}Limitations of the proposed model: given the
completely different style, the model cannot reproduce shading and
decoration. Example given by AB}

\end{figure}

Currently, the model has only been tested on protohistoric Italian
pottery, which means that the model cannot correctly handle painted
decoration that is not attested in the training dataset. Furthermore,
the model may have difficulty with drawing styles that are very
different from those in the training dataset. In this respect, the
relationship between decorations and shading is particularly important.
A closer look at Figure~\ref{fig-limit_1}, reveals that the model used
(\emph{6h-MCG}) has succeeded in correctly rendering all the structural
elements of the vessel (outline, fractures and handle), but has failed
to distinguish between decorations and shading, creating a `dirty'
effect on the drawing that is not aesthetically pleasing and does not
allow the decorations to be distinguished. This is a problem that can
easily be solved by a small fine-tuning of the model, letting it learn
the difference between decorations and shading.

Regarding scalability, such as complex image translation needs a
powerful hardware. DDPMs are indeed computationally intensive.
Archaeologists usually work with photogrammetry, GIS environments and
general-purpose graphic. In this sense, a machine used for this kind of
work should be able to handle the pipeline. A different approach needs
the training or the fine-tuning of the model. In this scenario, a
server-like of high-end gaming GPU is needed, as the training process
needs \textgreater{} 20 GB of VRAM. This is the main limitation of the
proposed method, as not all researchers (and especially archaeologist)
have direct access to such hardware. I cannot provide a solution to this
problem, as it is a limitation of the DDPMs themselves: high-quality
results need resources. However, a possible solution is the use of cloud
services (for example, Google Colab), which provide access to powerful
GPUs for a fee.

\subsection{Ethical Considerations}\label{ethical-considerations}

I obtained explicit permissions from the original drawers for using
their work in training this model. I acknowledge the potential impact of
automation on the professional practice of archaeological illustrators
and address these implications directly.

The application of generative AI in artistic creation remains an active
subject of debate (Amanbay 2023; Wang 2023; Zhou and Lee 2024). The
model specifically generates digitally inked archaeological drawings,
and a clear disclosure of AI assistance in any resulting illustrations
is mandatory. This transparency ensures that viewers can distinguish
between traditional and AI-assisted works, maintaining the scholarly
integrity of archaeological documentation.

The model is designed as a complementary tool rather than a replacement
for human expertise. It requires hand-drawn pencil sketches as input,
preserving the critical interpretative role of archaeological
illustrators. This dependency ensures that the foundational work - the
identification, interpretation, and initial documentation of
archaeological artifacts - remains firmly in human hands. The model
simply automates the inking process, a traditionally time-consuming but
technically straightforward task. The model's output can be further
edited or supplemented manually, allowing illustrators to add detail,
correct errors, or enhance the drawing as needed.

\section{Conclusion and Future Work}\label{conclusion-and-future-work}

\emph{PyPotteryInk} represents a significant advancement in
archaeological documentation, automating the time-consuming process of
inking pencil sketches to produce publication-ready drawings. In this
way, the model helps to increase the amount of data available for
research. The results of the model training and expert evaluation
demonstrate its effectiveness in generating high-quality illustrations
suitable for academic publication. For archaeologists, this tool offers
a substantial reduction in processing time, enabling the rapid
production of publication-grade drawings for multiple artifacts. Within
the future work, the model needs to be applied to other class of
materials, like lithics or metal objects. The open-source nature of the
model and the project encourages the community to contribute to the
development of the model, and to adapt it to different contexts, styles
and materials.

\section{Acknowledgements}\label{sec-acknowledgements}

The author would like to thank the archaeological illustrators who
provided the drawings used in this study, as well as the experts who
participated in the evaluation (FE: Federico Erbetti; EF: Elena Fausti;
LP; Lucrezia Petrucci; CP: Costanza Paniccia; EP1: Elisa Paolini; EP2:
Elisa Pizzuti; ADR: Andrea Di Renzoni; AC: Andrea Cardarelli). The
author would also like to thank Andrea Cardarelli (chair of European
Protohistory, Sapienza University of Rome) and Gianluca Pellacani (Museo
Civico di Modena) for providing the drawings used in the training
dataset. Thanks also to AB (Alessia Bovio) for providing some of her
drawings.

\section*{References}\label{references}
\addcontentsline{toc}{section}{References}

\phantomsection\label{refs}

\end{document}